\documentclass[journal=jpca,manuscript=letter,layout=shared]{achemso}
\usepackage{graphicx}
\usepackage{color}
\usepackage{natbib}
\usepackage[version=3]{mhchem}
\usepackage{amssymb}
\usepackage{enumerate}

\author{Mathias Rapacioli}
\email{mathias.rapacioli@irsamc.ups-tlse.fr}
\affiliation{Laboratoire de Chimie et Physique Quantiques LCPQ/IRSAMC, UMR5626, Universit\'e de Toulouse (UPS) and CNRS, 118 Route de Narbonne, F-31062 Toulouse, France.}

\author{Nathalie Tarrat}
\email{nathalie.tarrat@cemes.fr}
\affiliation{CEMES, Universit\'e de Toulouse, CNRS, 29, rue Jeanne Marvig, 31055 Toulouse, France}

\author{Fernand Spiegelman }
\affiliation{Laboratoire de Chimie et Physique Quantiques LCPQ/IRSAMC, UMR5626, Universit\'e de Toulouse (UPS) and CNRS, 118 Route de Narbonne, F-31062 Toulouse, France.}

\title{Melting of the Au$_{20}$ gold cluster : does charge matter? }
\begin{tocentry}
\includegraphics[width=4cm]{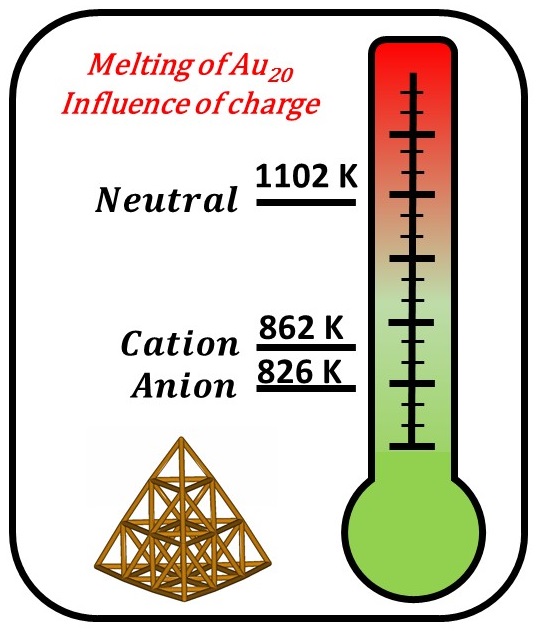}
\end{tocentry}

\begin{document}

\maketitle\begin{abstract}
We investigate the dependence upon charge of  the heat capacities  of the magic gold cluster Au$_{20}$ obtained from  density functional based tight binding theory within parallel tempering molecular dynamics and the multiple histogram method. The melting temperatures, determined from heat capacity curves, are found to be 1102 K for neutral Au$_{20}$ and only 866 and 826 K for Au$_{20}^+$ and Au$_{20}^-$. The present work proves that a single  charge quantitatively affects the thermal properties of the twentymer even for a global property such as melting.  
\end{abstract}

\section{Introduction}

Investigation of the thermodymical behaviour of small metal clusters has strongly developed since the early  investigations by Pauwlov\cite{Pawlow1909} and  Buffat and Borel\cite{buffat76} on  finite size gold  particles, showing that small particles have a lower melting point than bulk materials. After the pioneering  nanocalorimetry experiment of Schmidt {\it et al}.\cite{schmidt98}, 
a couple of experimental techniques\cite{breaux03,chirot08,boulon14} have made possible the determination of the heat capacity curves of metal clusters as a function of temperature down to selected sizes as small as typically ten atoms, in particular in the region of the  finite size equivalent of the solid-liquid transition. Unlike in the bulk, the transition in finite systems is not abrupt but extends over a finite temperature interval, as formalized by Berry\cite{Berry90} and documented in textbooks\cite{wales2004,labastie07}. Both the melting temperatures and  the latent heat have been  shown to depend strongly on size\cite{schmidt98}.  Alkali metal clusters have been  specially documented, however caloric curves of other  metal particles have also been experimentally obtained.  For other types of  clusters,  the influence of a single impurity\cite{douady09} or a substrate\cite{Loliveira2015} on the thermal properties has also been documented.\\       

Gold clusters and nanoparticles have been the object of numerous studies, due to their remarkable properties in several application fields such as  catalysis, nano-electronics, nano-luminescence or medicine. The structure and static properties of Au$_N$ clusters and nanoparticles have been widely investigated, some  of the works providing explicit comparison between  data from calculated structures  (mostly in the DFT framework) and experimental  data. 
However although  the early experiment of Buffat and Borel\cite{buffat76} were concerned with gold nanoparticles, no  experimental size-selected determination of the caloric curves in the cluster regime range (N less than 50) has been published  so far to our knowledge. A number of theoretical investigations\cite{castro90,lewis97, cleveland98,cleveland99,wang04,ruizgomez07, golovenko13,gafner15} usually  achieved with many body potentials or thermodynamical models have examined thermodynamically induced  structural conversion and melting  of large nanoparticles. 
Simulations of thermal properties were also carried out in the cluster regime size (from one to a few tens).  Several works were concerned by simulations of specific thermal behaviours related to  2D-3D transitions in small clusters\cite{koskinen07,li15} or to thermodynamical aspects of vibrational heating\cite{vishwanathan17}. The caloric curves of gold clusters in the small and medium size range were quite systematically investigated by Soul\'e du Bas and coworkers\cite{soule06,soule07}. In their comparison between  Au$_{19}$ and Au$_{20}$ in particular, they showed that strong differences could be induced by a single missing atom, which is a manifestation of finite size effects where each atom and defect counts. Differential effects were also observed in cage gold clusters.
Au$_{20}$ is referred to as a magic cluster.
The shape of the neutral cluster was found to be a  highly symmetric $T_d$ pyramid that can be viewed as a part of the $fcc$ lattice. The shapes of the ionized clusters Au$_{20}^+$ and Au$_{20}^-$ are also essentially pyramidal,   with very small Jahn-Teller deformations from the $T_d$  symmetry.  Those geometries are now well established from theory\cite{gruene08,letchke08} and   also from experiment, namely infrared  spectroscopy\cite{gruene08} for the neutral species, Trapped Ion Electron Diffraction \cite{letchke08} or Ion mobility\cite{furche02} techniques  for the cation and the anion. The  electronic structure of neutral Au$_{20}$ can be modeled as a closed shell system with $1s^21p^61d^{10}2s^2$ superorbital  configuration for the outer delocalized electrons in the simple spherical Jellium model (eventhough $5d$ electron bonding and atomic $6s-5d$ hybridization  can certainly not be neglected). This electronic shell closing together with the  symmetric $fcc$  packing of Au$_{20}$ supports a particular large stability and its magic character in the mass spectra. \\  

In recent works, we have adapted and  benchmarked DFTB parameters\cite{koskinen06,Fihey2015} for gold materials from clusters up to bulk \cite{agAu}. We additionally checked the convergence of the cohesive energies of larger nanoparticles of a few hundreds of atoms to the bulk values, as well as the structural, elastic  and energetical properties of bulk itself. As a key advantage, DFTB proved able to yield differential and selective results for charged clusters, namely Au$_n^+$ and Au$_n^-$ in addition to the neutrals, providing  fairly consistent ionization potentials  and  electron affinities. In a subsequent work, we combined DFTB with a  global search algorithm based on  the Parallel Tempering Molecular Dynamics  (PTMD) scheme  \cite{Sugita1999} completed with periodic quenching   to obtain the lowest energy isomers of Au$_{20}^{(0,+,-)}$ and Au$_{55}^{(0,+,-)}$.  One conclusion of the work for Au$_{20}$ was that the  isomerization energy gap is related to the specific charge state of the cluster, namely the gap is large for the neutral and much smaller for the anion and the cation. The present work is dedicated to the investigation of the influence of the charge state of the  cluster on the heat capacity curves in the temperature region about the  solid-liquid transition temperatures in the bulk.  Apart of this fundamental aspect, consideration of ions is also of interest because they can be more easily formed and detected in most experiments.
The nature of the phase changes in the interval 100-1700 K and  its dependence upon charge 
is  analysed from the caloric curves and the temperature-evolution of the isomer populations.

\section{Computational details} 

The potential energy of the neutral, cationic and anionic clusters were determined using the second-order version of DFTB \cite{dftb1,dftb2,scc-dftb} (Self-Consistent Charge SCC) and the parameters for gold
introduced in our previous publication\cite{agAu,Au55}. Note here that the SCC scheme is relevant  since it provides 
a self-consistent account  of electrostatics, especially important  at the surface of charged metal     clusters.
The Potential Energy Surface (PES) at various temperatures was explored using classical Molecular Dynamics with Parallel Tempering scheme  \cite{Sugita1999}, as previously implemented \cite{Oliveira2015} in the deMonNano code\cite{deMonNano}.  This technique strongly enhances the ergodicity in the simulations. For each case, we used a temperature range going from 50 to 2000 K  with 60 replica using an exponential distribution of temperatures. The length of each trajectory was 7.5 ns using a timestep of 1.5 fs to integrate the classical equations of motion. Exchanges were attempted using the Metropolis energy criterion every 1.2 ps. We used a Nos\'e-Hoover chain of thermostats with frequencies of 800 cm$^{-1}$ to achieve an exploration in the canonical ensemble. Evaporation did not occur as a crucial problem here, since there is a quite large difference between the solid-liquid and the liquid-gas  transition
temperatures (respectively 1337 K\cite{kaye} $vs$ 3243 K \cite{zhang}for the bulk). Nevertheless in order to  avoid any  problems in  the solid-liquid region of the heat capacity, we enclosed the clusters within a large rigid spherical potential centered on the cluster center of mass defined by  $V (r) = a(r -r_0)^8 $  with a=0.08 Hartree and $r_0$= 20 \AA~ for $r>r_0$.
Several  approaches exist to determine the heat capacity curve. In the present study, as in the work of Krishnamurty {\it al}.\cite{soule07}, we applied the multiple histogram method developed by Labastie and Whetten\cite{labastie90}. This approach reduces the statistical noise and allows to extrapolate heat capacities to temperatures not explicitly simulated. \\

  Finally, the isomer population analysis was done after periodic quenching. For each temperature, 1024 quenches (conjugated gradient optimization), 
regularly distributed along the trajectory
were achieved, allowing to assign the isomer basins (catching areas).  Identification of the quenched structures  was done combining an energy threshold (difference in energy less than 10$^{-4}$ Hartree) together with  an ordered-distances criterion (the interatomic distances based similarity function introduced by Joswig {\it et al}.\cite{simil} with a similarity threshold of 0.95). Note that a number of extra  quasi-degenerate isomers, similar however not identical, were obtained with respect to our previous work\cite{Au55}. 

\section{Results and Discussion}

\begin{figure}
\includegraphics[width=8cm]{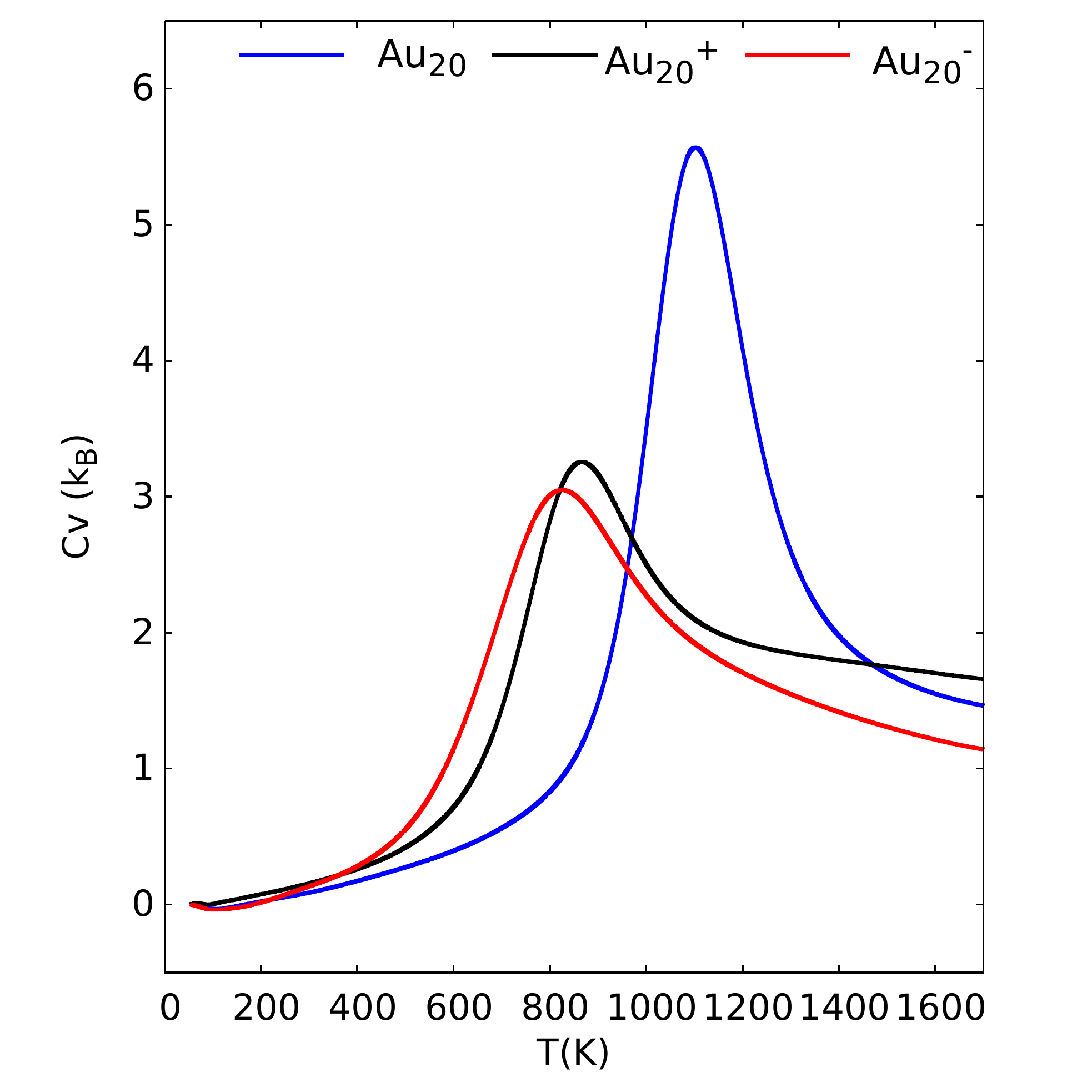}
\caption{Heat capacity curves for Au$_{20}$, Au$_{20}^+$ and Au$_{20}^-$.}
\label{fig:capa}
\end{figure}

	The heat capacity curves are plotted in figure \ref{fig:capa}. The first noticeable feature is that all three curves are unimodal with a single and well defined peak and that no premelting feature can be inferred from  the heat capacities (unlike for instance in Au$_{17}$\cite{chandrachud09}). Since the transition ranges are finite, we will comment here the temperatures $T_{start}$ at which the transition starts and the temperatures of the peak maxima which will supposedly be considered as the melting temperatures $T_{melt}$.  
One may estimate that the  transition starts at the intersection between the low temperature linear raise and the tangent at the inflexion point of the heat capacity. For the neutral cluster, T$_{start}$   is around  904 K  while this non 
linear raise occurs at lower temperature for  the cation and the anion, namely  642  and 551 K. The peak maxima for the neutral, the cation and the anion are located at  1102, 866 and 826 K respectively. Thus the temperature extension range of the transition can be estimated as 396, 448 and 550 K  respectively (estimated as the  peak bottom width taken as twice the difference between the start and the peak maximum). The present result for the neutral is in qualitative good correspondence with the DFT/LDA work of  Krishnamurty {\it et al.}\cite{soule07} who also observed a steep raise in the temperature evolution of the root mean square displacement attributed to a direct solid-liquid transition for  Au$_{20}$. Note that the peak maximum in the present work is shifted to significantly higher temperatures, likely due to the larger isomerization energies in DFTB $vs$ LDA.  Let us mention  that the DFT isomerization energies may significantly depend on the functional used. Actually, the present DFTB first gap  to isomerization (0.64 eV) is in much better correspondence   with the DFT/TPSS gap  of Letchken {\it et al.}\cite{letchke08} (0.75 eV) than with the DFT/LDA results (0.44 eV). The width of the phase change interval ($\sim$400 K) in the present DFTB simulation is nevertheless  consistent with the heat capacity data  of  Krishnamurty {\it et al.} \cite{soule07}

\begin{figure*}
\includegraphics[width=16cm]{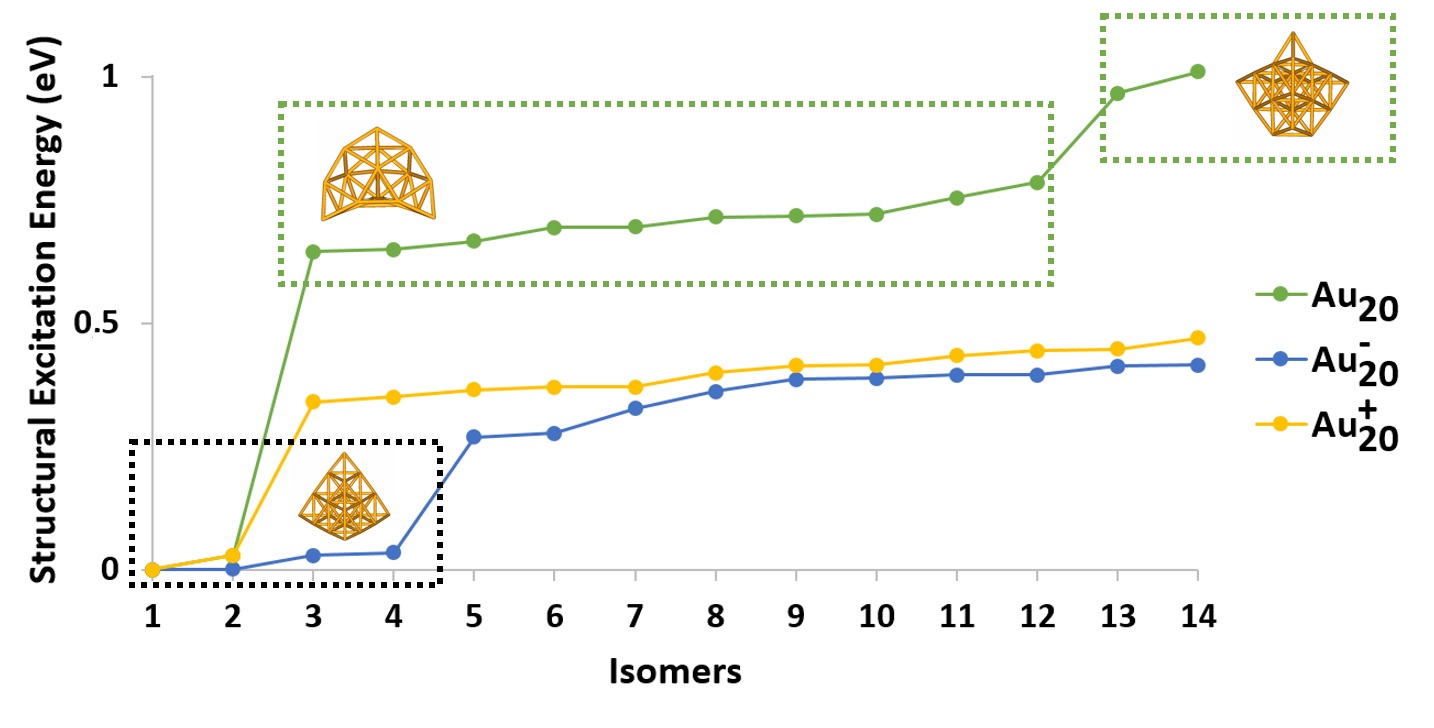}
\caption{Structural excitation energies of the 14th lowest energy isomers of Au$_{20}$ (green), Au$_{20}^+$ (yellow) and Au$_{20}^-$ (blue). Isomers belonging to the same meta-basin  are framed using boxes (see supporting information). \label{fig:geo}}
\end{figure*}

The original outcome of the present work  concerns ions and shows that  there is a significant  reduction  of the melting temperature for charged clusters with identical  size, and even a perceivable  difference can be made between the cation and the anion. The DFTB isomers structural excitation energies  are reported in fig~\ref{fig:geo}. Note that the quenching procedure used here (achieving quenches from all the MDPT trajectories at all the samples temperatures) generated extra isomers in addition to those of our previous work among the 15 lowest ones for each charge state. As mentioned above, the lowest energy  isomer of all three species  is  a pyramidal-like $fcc$   structure. One may also mention that quasi-degenerate isomers with this lowest energy structure are obtained consisting of very small distortions. In the ions, these correspond to various Jahn-Teller distortions, generating several, almost isoenergetics isomers. The above deformations  only affect the T=0 K limit of the heat capacities, however they do not create any visible associated feature on the heat capacity curves.  Note that the T=0 K limit  cannot be properly described in the present work due to the neglect of quantum effects. Let us however note that  several groups of quasi-degenerate isomers with very neighboring topologies can be found below 1 eV. For instance, isomers 3-12 of the neutral are distinct but they only differ by one or two bond or a small distortion (see Supplementary Information figure 2). Such sets of isomers may be thought of as defining  meta-basins of the PES. The main differences between the neutral, the cations and the anions however concern the  first non-pyramidal isomers. In Au$_{20}$, the set of isomers 3-12  lie in the range 0.64-0.79 eV above the lowest one and  the isomerization energy of the next one (13) is 0.97 eV, not considering the barriers.  In the cation, the   isomerisation energy to isomer 4  is 0.34 eV. In the anion, the isomerization energy to isomer 5 is  0.27 eV. Both are significantly smaller than the corresponding gap for the neutral. A continuous progression above the first gap is observed. Again this decrease of the structural excitation energies above the pyramid meta-basin  for ions with respect to the neutral are in reasonable correspondence with the DFT/TPSS values of Lechtken\cite{letchke08}.

\begin{figure}
\begin{tabular}{c}
\includegraphics[width=6.2cm]{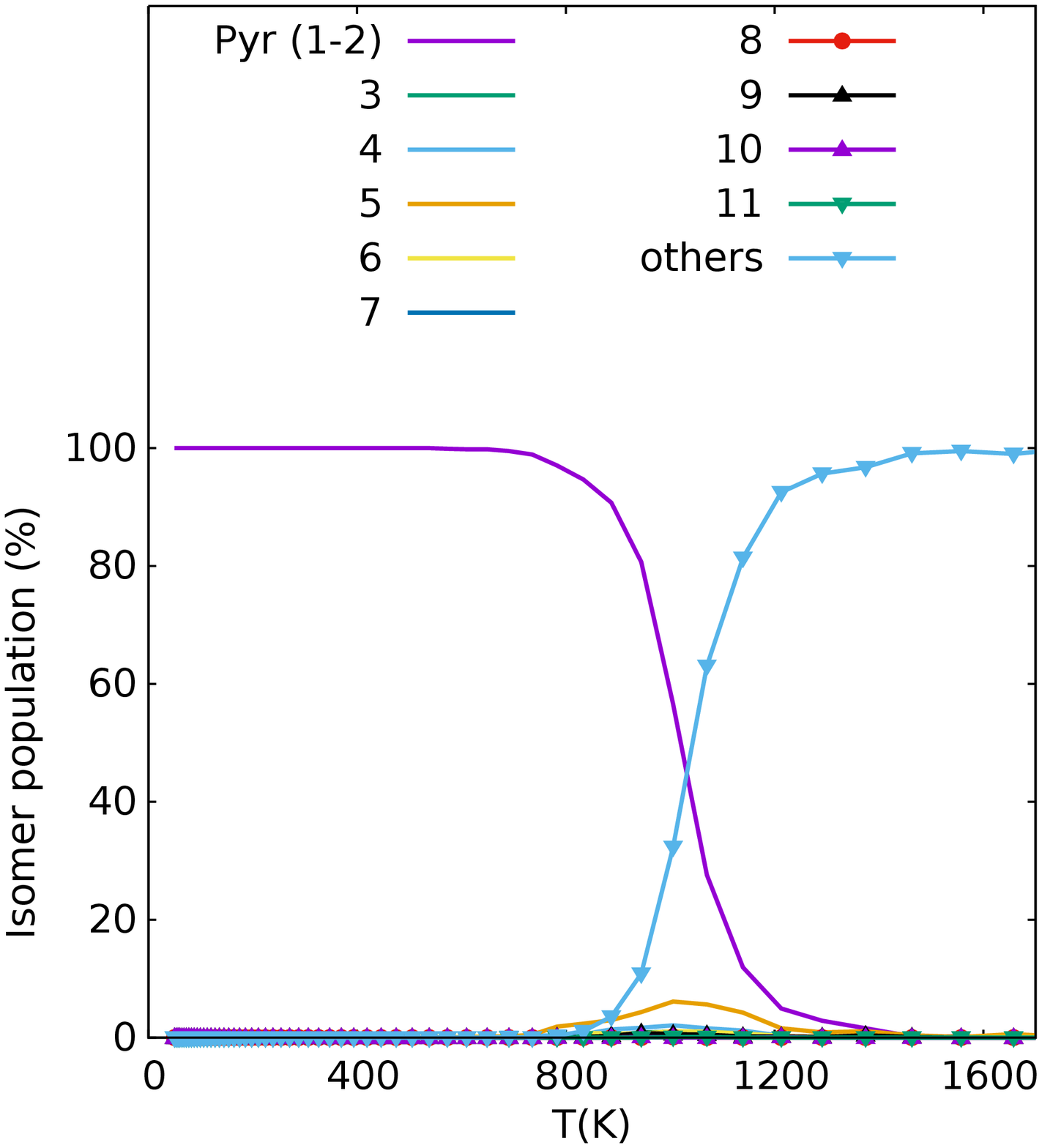} \\
\includegraphics[width=6.2cm]{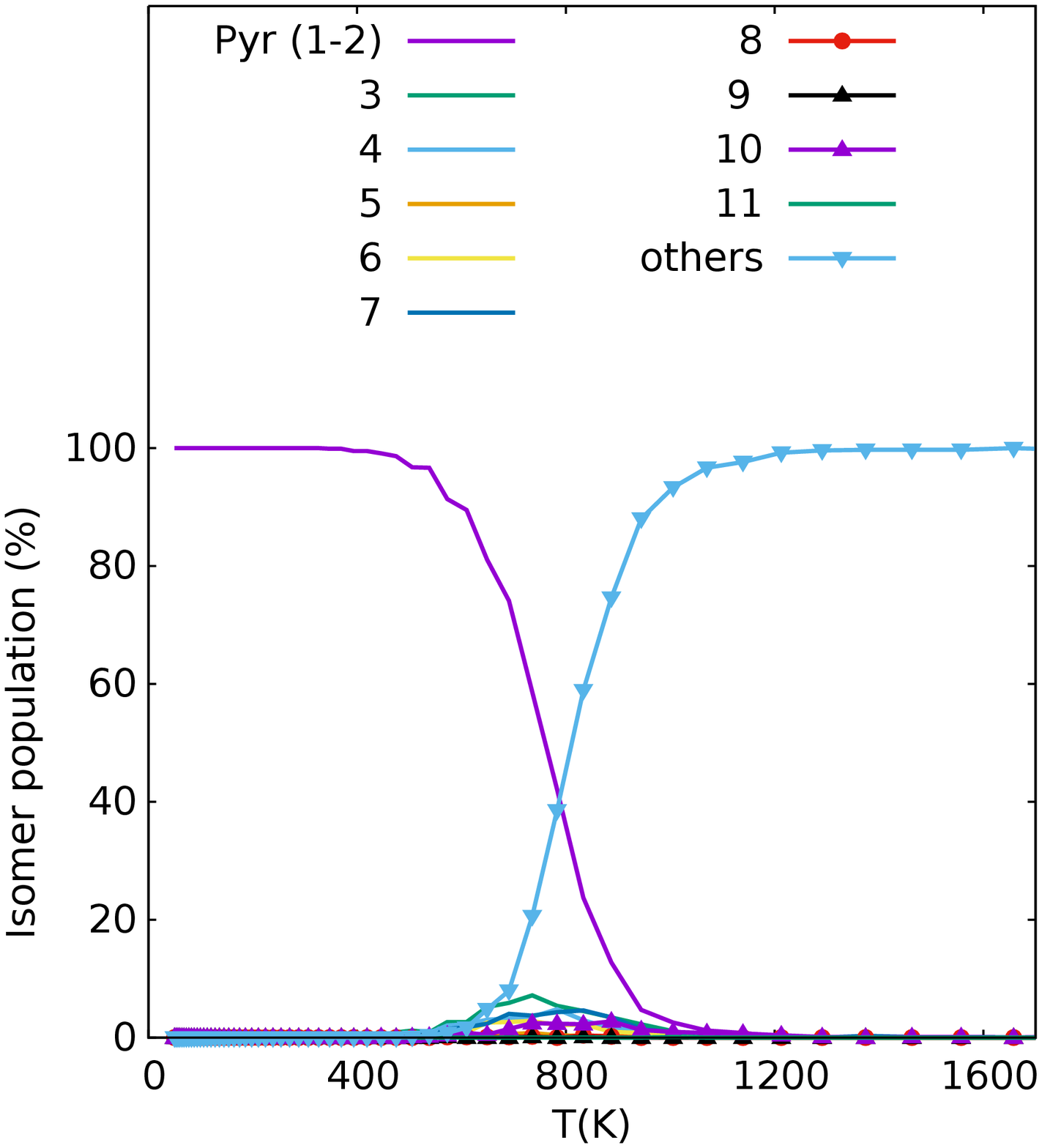}\\
\includegraphics[width=6.2cm]{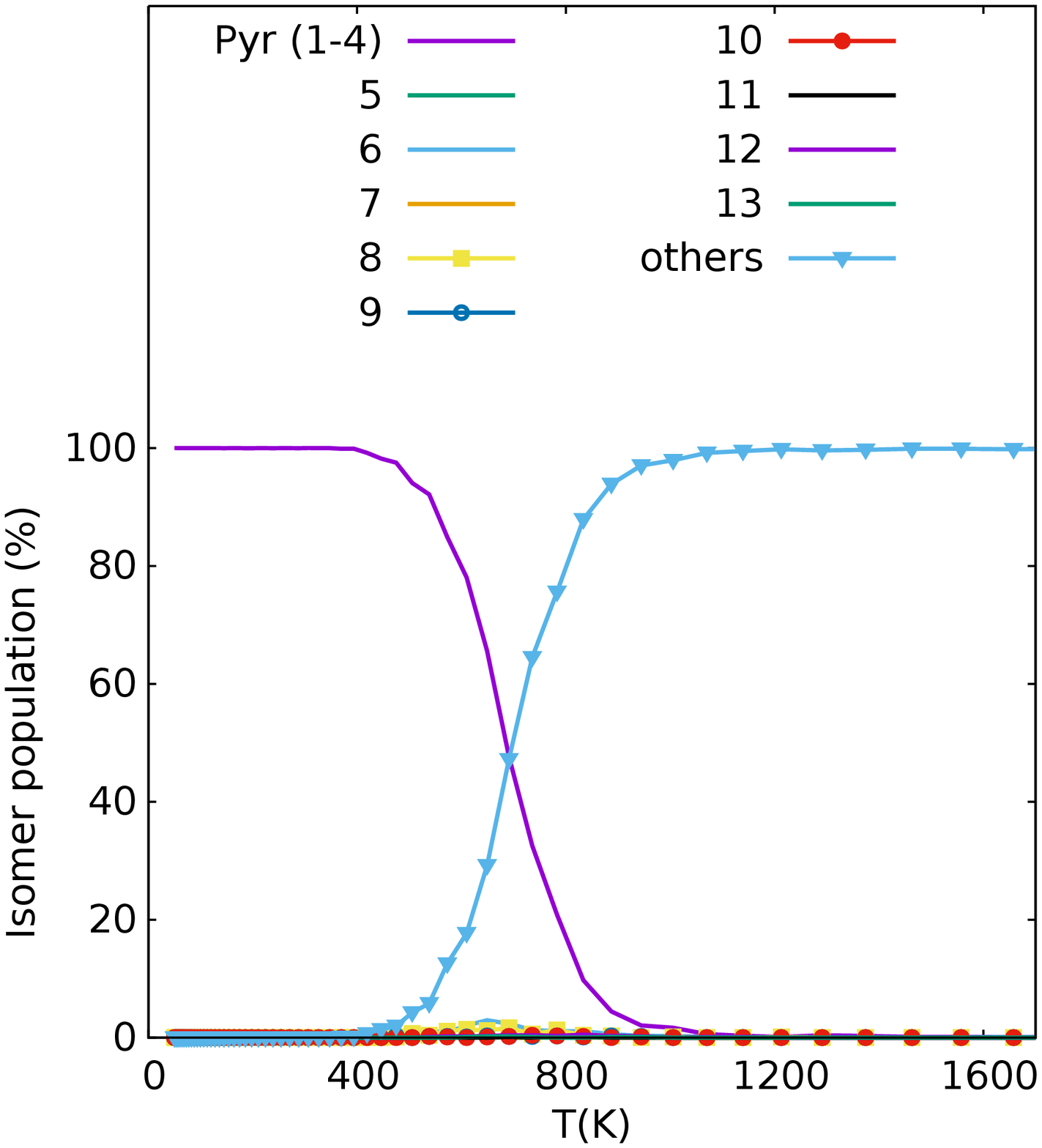}
\end{tabular}
\caption{Temperature evolution of the isomer population as a function of Temperature for Au$_{20}$ (top), Au$_{20}^{+}$ (center) and Au$_{20}^{-}$ (bottom).}
\label{fig:pop}
\end{figure}

 Insight in the atomistic aspect of melting can be gained in the analysis of the isomer basin populations.  
In the neutral case, it appears that the
decrease of the population of the pyramidal isomer starts around 800 K. The decrease is quite regular, not showing any steps,  and the depletion is fully achieved around 1300 K. Only a single meta-basin (isomers in the range 0.64-0.76 eV, essentially isomer 5) contributes  with a visible population (however minor $<$ 10 percent at its maximum)  within the melting temperature range, namely 700-1200K and then disappears at higher temperature.   The main observation for T $\geq$ 900 K, is a growing population consisting of a collection of many different higher energy  isomers (labeled "others" in fig~\ref{fig:pop}),  revealing strong geometrical fluctuations, so that the cluster can be considered as melted.  The  minor contribution  of some  meta-basin structures 3-12 thus  appears more as a prelude of a direct and continuous solid-to-liquid transition than as a solid-to-solid transition.  Almost all the lowest isomers are essentially generated by the migration of a corner atom (possibly two ) to a face of the pyramidal cluster, inducing  possible distortions.  Those results are qualitatively    consistent   with the conclusion of the DFT/LDA simulation of  Krishnamurty {\it et al.} \cite{soule07}, despite of the quantitative differences in  the melting temperatures between  the  two calculations.\\ 

The melting mechanisms corresponding to ionized species are similar. For the cation case, depletion of  the pyramidal lowest energy structure starts  as soon as 450-500K  while populations of isomers 3,4,7 and above somewhat  increase up to a maximum. At higher temperature, these population  decrease, while isomers labeled "others" become predominant as in the neutral case. In the anion none of the low-energy isomers  is ever significantly populated. Thus as for the neutral clusters, melting appears as a direct and continuous solid-liquid transition, however shifted to lower  temperatures,  due to the lowest energy needed for the departure from the pyramid basins. In no cases any onset  of a solid-to-solid type premelting feature can be significantly observed.

\section{Conclusion}
 
 Using the DFTB approach and canonical PTMD exploration scheme, we have  achieved a comparative analysis of the  heat capacity curves of Au$_{20}$, Au$_{20}^+$ and Au$_{20}^-$
 as a function of temperature.  The melting temperatures are estimated at  1102  K, 866 K and 826 K 
respectively. The present work shows that the change in the potential energy landscape induced by the  
cluster  charge yields   a significant  variation in  the energetical distribution of the low energy isomers and results in a quantitatively  different solid-to-liquid transition temperature. Nevertheless, in all cases the transition is found to correspond to direct melting. This extends the findings of Krishnamurty {\it et al.} \cite{soule07} addressing thermodynamical  finite size effects for gold. Although this difference will evidently vanish for large nanoparticles, the present work evidences charge  effects in clusters, namely the influence of an extra charge, even unity. This was already known for electronic stability for instance but is evidenced here also for a thermal global  quantity such as the heat capacity in  this medium size range, which still is shown to be  strongly determined not only by the  atom count, but also by the electron count.

\section{Acknowledgements}
This work was granted access to the HPC resources of CALMIP (Grants p1303 and p0059) and from IDRIS (Grant i2015087375). It was supported by a CNRS-Inphyniti Grant (ATHENA project), the CNRS-GDR EMIE and the NEXT grants  ANR-10-LABX-0037 in the framework of the {\it Programme des Investissements d$'$Avenir} (CIM3 and EXTAS projects).

\begin{suppinfo}
Structures of isomers 3 to 14 of Au$_{20}$. 

\end{suppinfo}

\bibliography{capa_Au}

\end{document}